\documentclass[aps,showpacs, nofootinbib]{revtex4}

\newcommand{\be}{\begin{equation}}
\newcommand{\ee}{\end{equation}}
\newcommand{\ben}{\begin{eqnarray}}
\newcommand{\een}{\end{eqnarray}}

\newcommand{\la}{{\lambda}}

\newcommand{\si}{{\sigma}}

\newcommand{\cO}{{\cal O}}

\newcommand{\cL}{{\cal L}}

\newcommand{\cS}{{\cal S}}
\newcommand{\cE}{{\cal E}}
\newcommand{\cB}{{\cal B}}
\newcommand{\cM}{{\cal M}}

\newcommand{\p}{\partial}
\newcommand{\na}{\nabla}

\newcommand{\tpsi}{\tilde \psi}

\newcommand{\tcF}{{\tilde {\cal F}}}
\newcommand{\tcA}{{\tilde {\cal A}}}

\newcommand{\tF}{\tilde F}
\newcommand{\tE}{\tilde E}
\newcommand{\tB}{\tilde B}
\newcommand{\tJ}{\tilde J}
\newcommand{\tA}{\tilde A}
\newcommand{\tQ}{{\tilde Q}}

\newcommand{\tn}{\tilde n}

\newcommand{\tla}{\tilde {\lambda}}

\newcommand{\ep}{\epsilon}

\newcommand{\bla}{\bar \lambda}

\newcommand{\ga}{\gamma}

\pacs{04.20.-q,~04.20.Cv,~04.50.Gh}

\begin{document}

\title{Mass angular momentum and charge inequalities for black holes in Einstein-Maxwell-axion-dilaton gravity }

\author{Marek Rogatko}
\affiliation{Institute of Physics \protect \\
Maria Curie-Sklodowska University \protect \\
20-031 Lublin, pl.~Marii Curie-Sklodowskiej 1, Poland \protect \\
marek.rogatko@poczta.umcs.lublin.pl \protect \\
rogat@kft.umcs.lublin.pl}

\date{\today}

\begin{abstract}
Mass angular momentum and charge inequalities for axisymmetric maximal time-symmetric initial data
invariant under an action of U(1) group, in Einstein-Maxwell-axion-dilaton gravity
being the low-energy limit of the heterotic string theory, is established. We assume that data set
with two asymptotically flat regions is given on smooth simply connected manifold. 
We also pay attention to the area momentum charge inequalities for a closed orientable
two-dimensional spacelike surface embedded in the spacetime of the considered theory.
\end{abstract}

\maketitle

\section{Introduction}
Description of the gravitational collapse dynamics is a real challenge to theoretical
investigations in the realm of Einstein gravity and its generalizations. Recently, one can observe
a big resurgence of the aforementioned problems originated from 
the researches conducted in Ref.\cite{pen73}. 
Revisiting original Brill's proof of positive mass \cite{bri59}, Riemaniann Penrose inequality was proved
in four and generalized to five-dimensional case of Einstein theory of gravity, in the context
of time-symmetric, axisymmetric initial data \cite{gib06}.
The total
mass angular momentum inequality, binding global quantities has been expanded to the dynamical
case of vacuum and electrovacuum axisymmetric spacetimes \cite{dai06}-\cite{lop10}. Further
perspicacity into the investigations in question was taking into considerations quasi-local quantities
characterizing black holes. The aforementioned inequalities were studied in axisymmetric spacetime
with matter besieged the event horizon \cite{ansorg}. The vacuum Einstein gravity case was treated in Refs.
\cite{dai10}-\cite{ace11}, where an inequality between area of the apparent horizon and angular momentum for a class of
axially symmetric black holes including initial conditions with isometry leaving fixed two-surfaces was conceived.
The initial data set of Einstein vacuum equations with cosmological constant was treated in\cite{dai11}.
On the other hand, the extension in order to incorporate electric and magnetic charges into the problem in question 
was elaborate in \cite{jar11}-\cite{cle12}. See also \cite{dai12} and references therein, for the recent review
of the problem and main ideas standing behind the proofs. The lower bound for a single black hole
in Einstein-Maxwell theory with axially symmetric maximal initial data and non-electromagnetic
matter fields satisfying dominant energy conditions was found in Ref.\cite{dai13} (see also \cite{mar12}). This 
inequality is
saturated only for the case when the initial data arise from extreme Kerr-Newmann spacetime.
\par
A natural extension of the predicament in question is
related to the problem of
gravitational collapse in generalization of Einstein theory to higher dimensions and
emergence of higher dimensional black objects. The complete classification of $n$-dimensional 
charged black holes both with non-degenerate and degenerate component of the event horizon was 
proposed in Refs.\cite{nd} but there were only partial results for the highly
nontrivial case of $n$-dimensional rotating black hole uniqueness theorem \cite{nrot}. These researches 
encompasses also the case of the low-energy limit 
of the string theory, like dilaton gravity, Einstein-Maxwell-axion-dilaton (EMAD)-gravity and 
supergravities theories \cite{sugra}. On the other hand, the strictly stationary static vacuum 
spacetimes in Einstein-Gauss-Bonnet theory were
discussed in \cite{shi13a}, while in Ref.\cite{shi12}
it was revealed that strictly stationary AdS spacetime could not allow for the existence of nontrivial
configurations of complex scalar fields or form fields. The generalization of the aforementioned problem,
i.e., strictly stationarity of spacetimes with complex scalar fields
in EMAD-gravity with negative cosmological constant was given in \cite{bak13}.
In Ref. \cite{shi12} it was revealed that a static asymptotically flat black hole
solution is unique to be Schwarzschild spacetime in Chern-Simons modified gravity. Then,
the uniqueness proof of static
asymptotically flat electrically charged black hole in Chern-Simons modified gravity was provided
\cite{rog13}. 
\par
Just, the inequalities between area and angular momentum 
in higher dimensional axisymmetric spacetime were given in \cite{hol12},whereas
inequalities binding area angular momentum 
and charges in Einstein-dilaton gravity were proposed in Ref.\cite{yaz13}. The five-dimensional extension
of the dilaton gravity was elaborated in \cite{yaz13b}. One should also
mention \cite{faj13}, where the inequalities for stable marginally outer trapped surfaces in dilaton gravity
were derived.
\par
Motivated by the aforementioned researches we shall search for the lower bound for the area of black
holes in EMAD-gravity being the low-energy limit of heterotic string compactified to four-dimensions.
One will not restrict himself to only one gauge field and take into account the 
arbitrary number of $U(1)$-gauge fields.
\par
Our paper is organized as follows. 
In Sec.II we present the underlying theory, defining the $SL(2,~R)$-duals to the gauge fields and
complex scalar {\it axi-dilaton}. Next, in Sec.III we 
find the general form of the total angular momentum and twist potential in the EMAD-gravity.
We find inequality binding angular momentum and {\it dilaton-electric} 
and {\it dilaton-magnetic} charges 
for a black hole with axially symmetric maximal initial data as well as
non-electromagnetic fields fulfilling the dominant energy condition.
Sec.IV will be devoted to the area angular momentum charge inequalities for a closed orientable
two-dimensional spacelike surface in manifold under consideration.

\section{EMAD-gravity}
Motivated by the recent works 
connected with inequalities binding black hole mass and other its parameters in Einstein-Maxwell (EM)
theory \cite{dai13}, we shall pose a question about such kind of inequalities in generalized
theory of gravity. 
Namely, in this section, we consider the so-called Einstein-Maxwell-axion-dilaton 
gravity (EMAD). The theory under consideration will contain gravitation 
field $g_{\mu \nu}$, arbitrary number $N$ of
$U(1)$-gauge fields, the dilaton field $\phi$ and axion $a$. 
The action for EMAD-gravity will be subject to the relation \cite{kal}
\be \label{emad}
S = \int d^4 x~{\sqrt{-g} \over 16\pi }~\bigg[
R  
- 2~\na_\mu \phi \na^\mu \phi - {1 \over 2}~\na_\mu a~\na^\mu a -
\sum \limits_{n=1}^N e^{-2 \phi}~F_{\mu \nu}^{(n)}~F^{\mu \nu ~(n)}
- \sum \limits_{n=1}^N a~F_{\mu \nu}^{(n)}~\ast F^{\mu \nu ~(n)}
\bigg],
\ee
where we have denoted the strength of the adequate 
gauge field $F_{\mu \nu}^{(n)} = 2~\na_{[ \mu}A_{\nu ]}^{(n)}$. On the other hand, its dual
is given by 
$\ast~F_{\mu \nu}^{(n)} = {1 \over 2}~\ep_{\mu \nu \alpha \beta}F^{\mu \nu~ (n)}$.
It should be remarked that when the number of vector fields is six we obtain $N=4,~d=4$ bosonic part of supergravity
theory.
In what follows, for the sake of generality, one will keep the arbitrary number of $U(1)$-gauge fields.
\par
It turned out that in many physical problems \cite{kal} the action describing by the relation (\ref{emad}) can be written
in a more convenient form.
Namely, introducing a complex scalar {\it axi-dilaton} in the form as
\be
\la = a + i~e^{-2 \phi},
\ee
and defining $SL(2,~R)$-duals to the gauge fields $F_{\mu \nu}^{(n)}$, the action in question 
implies
\be
S = \int d^4 x~{\sqrt{-g} \over 16\pi }~\bigg[
R + 2~{\na_\mu \la~\na^\mu \tla \over (\la -\tla)^2}
+ \sum\limits_{n=1}^N F_{\mu \nu}^{(n)}~\ast \tcF^{\mu \nu~(n)}
\bigg],
\label{ddu}
\ee
where the $SL(2,~R)$-duals are given by the relation
\be
\tcF_{\alpha \beta}^{(n)} = e^{-2 \phi}~\ast F_{\alpha \beta}^{(n)} - a~F_{\alpha \beta}^{(n)}.
\ee
The equation of motion for $SL(2,~R)$-duals is of the form
$\na_\alpha~\ast \tcF^{\alpha \beta~(n)} = 0$ and entails the existence of $N$ vector potentials
$\tcA_{\beta}^{(n)}$ satisfying relation
\be
\tcF_{\alpha \beta}^{(n)} = 2~\na_{[ \alpha} \tcA^{(n)}_{\beta ]}.
\ee 
Consequently, the analogous relation for $F_{\mu \nu}^{(n)} = 2~\na_{[ \mu} A^{(n)}_{\nu ]}$
is not a consequence of equations of motion but it stems from the Bianchi identity.
The energy momentum tensor for the complex scalar field and $U(1)$-gauge fields 
is provided by the following expression:
\be
T_{\alpha \beta}(F,~\tF,~\la) =
- \bigg[ 4~\sum\limits_{n=1}^N F^{(n)}_{\alpha \delta}~\ast \tcF_{\beta}{}^{\delta~(n)}
- g_{\alpha \beta}~\sum\limits_{n=1}^N F_{\alpha \beta}^{(n)}~\ast \tcF^{\alpha \beta ~(n)} \bigg]
+
{2~g_{\alpha \beta}~\na_{\ga}\la \na^\ga \bla - 4~\na_\alpha \la \na_\beta \bla \over
(\la - \bla)^2}.
\ee
\section{Mass inequalities for black hole in EMAD-gravity}
First 
we comment on the initial value formulation of EMAD-gravity equations with matter sources.
One assumes further that we have to do with non-electromagnetic matter fields.
We foliate the globally hyperbolic spacetime
by Cauchy surfaces $\Sigma_t$, which are parameterized by a global time $t$. Let $n_\alpha$
be the unit normal to the aforementioned hypersurface, then $n_{\alpha}~n^\alpha = - 1$.
Just the spacetime metric induced on $\Sigma_t$ a spatial metric $h_{\alpha \beta}$, by the relation
\be
h_{\alpha \beta} = g_{\alpha \beta} + n_\alpha~n_\beta.
\ee
On this account,
we define {\it electric} and {\it magnetic} components for gauge field strengths $F_{\alpha \beta}^{(n)}$
and $\tcF_{\alpha \beta}^{(n)}$. Namely, {\it electric} components imply
\be
E^{\alpha ~(n)} = - F^{\beta \alpha (n)}~n_\beta, \qquad
\tE^{\alpha ~ (n)} = - \tcF_{\beta \alpha (n)}~n_\beta,
\ee
while {\it magnetic} ones are provided by the following relations:
\be
B^{\alpha ~(n)} = - \ast F^{\ga \alpha~(n)}~n_\ga, \qquad
\tB^{\alpha ~ (n)} = - \ast \tcF^{\ga \alpha~(n)}~n_\ga,
\ee
where one denotes, respectively
\ben
\ast F^{\ga \delta~(n)} &=& {1 \over 2}~\ep_{\alpha \beta \ga \delta}~k^\beta~F^{\ga \delta~(n)},\\
\ast \tcF^{\ga \delta~(n)} &=&
{1 \over 2}~\ep_{\alpha \beta \ga \delta}~k^\beta~\tcF^{\ga \delta~(n)}.
\een
A complete initial data for the theory provided by the action (\ref{ddu}) will consist of the initial
Cauchy hypersurface $\Sigma_t$, induced metric on it, its extrinsic curvature 
$K_{ij}$, the value
of {\it axi-dilaton} complex scalar $\la$ and $n^j~D_j \la$ on $\Sigma_t$ and 
{\it electric} and {\it magnetic} fields for each of the n-th gauge components defined on the hypersurface 
in question. Moreover, if $\Sigma_t$ is time symmetric one has that $n^j~D_j \la = 0$.\\
By virtue of the above definitions and properties of the {\it electric} and {\it magnetic}
components of the adequate gauge strength field $F_{\alpha \beta}^{(n)}$
and $\tcF_{\alpha \beta}^{(n)}$, one concludes that 
the constraint equations for EMAD-gravity are provided by 
\ben \label{div}
D^a~\bigg( K_{ab} &-& K_c{}{}^c~h_{ab} \bigg) - 2~
\sum\limits_{n=1}^N
\ep_{bdj}~B^{(n)j}~\tB^{(n)d} = 8\pi~P_b, \\ \label{kk}
{}^{(3)}R + (K_a{}{}^a)^2 &-& K_{ij}~K^{ij} - 2~\bigg[
\sum\limits_{n=1}^N
\bigg( B_i^{(n)}~\tE^{(n)i} - E_j^{(n)}~\tB^{(n)j} \bigg) - {\chi_a~{\tilde \chi}^a \over (\la - \bla)^2} \bigg]
= 16\pi~\mu,
\een
where $\chi_a = D_a \la$, $D_a$ is the derivate with respect to $h_{ab}$ metric while
$P_b$ matter momentum density and $\mu$ is matter energy density.
In our considerations we assume that matter fields will satisfy the dominant energy condition
$\mu \ge \mid P_i \mid$. Thus, equations (\ref{div}) and (\ref{kk}) define the time symmetric
initial data for the theory under consideration.
\par
In what follows we shall consider asymptotically flat Riemaniann manifold in which there exists a region
diffeomorphic to $R^3 \setminus B(R)$, where $ B(R)$ is a coordinate ball of radius $R$. In local coordinates
on the above region the adequate fall-off conditions are required to satisfy
\ben
h_{ij} &-& \delta_{ij} = \cO_k(r^{-{1 \over 2}}), \qquad
\p_k h_{ij} \in L^2(M_{ext}), \qquad K_{ij} = \cO_{l-1}(r^{-3}), \\ \nonumber
E^i &=& \cO_{l-1}(r^{-2}), \qquad
\tE^i = \cO_{l-1}(r^{-2}), \\
B^i &=& \cO_{l-1}(r^{-2}), \qquad
\tB^i = \cO_{l-1}(r^{-2}),
\een
where we have denoted $f = \cO_k(r^\la), ~\p_{k_1 \dots k_l} f = \cO(r^{\la -l})$, for
$0 \leq l \leq k$. 
\par
Next, one commences with the initial data set for EMAD-gravity consisting of metric tensor,
extrinsic curvature and vector connected with the gauge fields in the underlying theory, i.e.,
$(M,~h_{ij},~K_{ab},~E_i,~\tE_i,~B_a,~\tB_a,~\la)$. Furthermore, in Ref.\cite{chr08} it was revealed that
in the case of simply connectedness of the manifold in question, the analysis reduced to the considerations
of manifold $R^3 \setminus \sum\limits_{j=1}^H a_j$, where $a_j$ are points in $R^3$ representing
asymptotic ends. There also exists a global cylindrical Brill coordinate system, where $a_j$ lie
on $z$-axis. The fall-off conditions in asymptotically flat ends avouch the definitions of the ADM mass
and the adequate charges
\ben
m &=& {1 \over 16\pi} \int_{S_{r \rightarrow \infty}} dS~\bigg( h_{ij,i} - h_{ii,j} \bigg)~\tn^j,\\
Q_e^{(n)} &=& {1 \over 4\pi}\int_{S_{r \rightarrow \infty}} dS~E_a~\tn^a, \qquad
Q_m^{(n)} = {1 \over 4\pi}\int_{S_{r \rightarrow \infty}} dS~B_a~\tn^a, \\
\tQ_e^{(n)} &=& {1 \over 4\pi}\int_{S_{r \rightarrow \infty}} dS~\tE_j~\tn^j, \qquad
\tQ_m^{(n)} = {1 \over 4\pi}\int_{S_{r \rightarrow \infty}} dS~\tB_j~\tn^j,
\een
allied with {\it electric} and {\it magnetic} components of the gauge strength fields 
$F_{\alpha \beta}^{(n)}$ and $\tcF_{\alpha \beta}^{(n)}$, respectively.
\par
Now we shall take into account the axisymmetric initial data, i.e., data that are invariant under the action
of $U(1)$ group. On this account axisymmetric feature is encoded in the line element of the form
\be
ds^2 = q_{AB}~dx^A dx^B + X^2~\bigg( d\varphi + W_B~dx^B \bigg)^2,
\ee
where 
$q_{AB}$ is a two-dimensional metric on the orbit space of Killing vector
$\eta_\alpha = ( \p/\p \varphi)_\alpha$ and moreover the functions $X$ and $W_B$ are independent on
$\varphi$-coordinate. It turns out that the {\it strongly axisymmetric} condition
input additional mirror symmetry and causes that $W_B$ has to disappear \cite{gib06}.\\ 
One can find such coordinate that
\be
ds^2 = e^{-2U + 2\alpha}~(d\rho^2 + dz^2) + \rho^2~e^{- 2U} ~\bigg(
d \varphi  + \rho~W_\rho~d\rho + W_z~dz\bigg)^2,
\ee
where all the functions are $\varphi$-independent. 
The above choice of the line element leads to finding a harmonic function on the orbit space,
i.e., $\na_i\na^i_{(q_{AB})} \rho = 0$ and specifying conditions at infinity and on the $z$-axis.
Moreover certain conditions on functions $U$ and $\alpha$ in order to obtain regularity
of the axisymmetric line element should be imposed \cite{chr08,gib06}.
\par
As we shall exploit the {\it axisymmetric initial data} which make the group of
manifold isometries include $U(1)$-subgroup, the defined quantities should be invariant under
the aforementioned group action. Namely, we have that
\be
\cL_\eta h_{ab} = \cL_\eta K_{ij} = \cL_\eta E_i^{(n)} = \cL_\eta \tE_i^{(n)} = \cL_\eta B_j^{(n)}
= \cL_\eta \tB_j^{(n)} = 0,
\label{sym}
\ee 
where $\cL_\eta$ is Lie derivative with respect to the Killing vector field $\eta_\alpha$.
In Ref.\cite{dai13} it was revealed that in EM-theory the angular momentum in the direction of the
rotation axis, of two-dimensional surface $\Sigma \in M$, with a tangent vector $\eta_\alpha$ and
$\tn_i$ unit outer normal over the coordinate sphere, can be written as
\be
J(\Sigma) = {1 \over 8\pi} \int_{\Sigma} d\Sigma~\bigg( K_{ij} - K_{a}^{a}~h_{ij} \bigg)~\tn^i~\eta^j.
\label{jj}
\ee
One should comment that equation (\ref{jj}) describes the {\it Komar-like} angular momentum
connected with a two-dimensional surface with the axial vector $\eta^i$, that coincides with the
Komar definition of angular momentum when $\eta^i$ can be expressed in the vicinity of $\Sigma$.\\
But $J(\Sigma)$ is not necessary conserved. 
The crucial point is that we consider the matter fields like $U(1)$-gauge fields, dilaton and axion fields,
which the standard bulk contribution may be written in terms of Stoke's theorem, using a surface term in a natural
way associated with black hole.
Hence, one is motivated to define the total angular momentum on a hypersurface $\Sigma$, with
contributions of gauge fields in the underlying theory, which has this property. Having in mind this idea, we postulate
the total angular momentum provided by the following expression:
\ben
\tJ(\Sigma) &=& {1 \over 8 \pi} \int_{\Sigma} d\Sigma~\bigg( K_{ij} - K_{a}^{a}~h_{ij} \bigg)~\tn^i~\eta^j \\ \nonumber
&+& {1 \over 4\pi} \sum\limits_{n=1}^N~\int_{\Sigma} d\Sigma~A_{k}^{(n)} \eta^k~\tn_i \tB^{i (n)}.
\een
The motivation for introducing the second term was mainly to obtain the conservation of the total
angular momentum. Moreover, one has that if we set dilaton and axion fields equal to zero and restrict considerations to
the only one gauge field, we arrive at the form of the potential in Einstein-Maxwell (EM) theory \cite{dai13}. 
On the other hand, the form depending on $\sum\limits_{n=1}^N~\tB^{i (n)}$ (not like in EM-theory on $E_j$) has its roots
in equation of motion for EMAD-gravity. Namely, in the theory under consideration one has that the divergence of
$\sum\limits_{n=1}^N~\tB^{i (n)}$ is equal to zero. Contrary to Maxwell electrodynamics when $\na_j E^j = 0$.
\par
Despite of the fact that the potentials $A_{j}^{(n)}$ are discontinuous on the $z$-axis, the product
$\sum\limits_{n=1}^N A_{k}^{(n)}~\eta^k$ remains well behaved, because of the fact that
the Killing vector field $\eta_\alpha$ vanish on the $z$-axis.
\par
Now we restrict our attention to the problem of the total angular momentum in the theory under
consideration. To proceed further, we
shall consider the second term on the right hand-side of relation (\ref{div}). It yields 
\ben
\ep_{kij}~\tB^i~B^j ~\eta^k &=& \ep_{kij}~\tB^i~\ep^{jab}~D_a A_b~\eta^k =
D_a \bigg( \ep_{kij}~\tB^i~\ep^{jab}~A_b~\eta^k \bigg) \\ \nonumber
&-& \ep_{kij}~D_a \tB^i~\ep^{jab}~A_b~\eta^k - \ep_{kij}~\tB^i~\ep^{jab} ~A_b~D_a \eta^k\\ \nonumber
&=& D_a \bigg( \ep_{kij}~\tB^i~\ep^{jab}~A_b~\eta^k \bigg) + A_k~\eta^k~D_j \tB^j,
\een
where we have used the invariance properties under the motion of $U(1)$ group. Consequently, let us 
take into account a domain of 
the manifold in question, $M_1 \in M$, with boundaries $\p M_1 = \Sigma_1 \cup \Sigma_2$
\be
\int_{M_1} dV~ \ep_{kij}~\tB^i~B^j ~\eta^k  = \int_{M_1} dV~A_k~\eta^k~D_j \tB^j -
\int_{\p M_1}d\Sigma~ A_k~\eta^k~\tB^j~\tn_j.
\ee
In derivation of the above equation one has to take into account that 
the Killing vector field
$\eta_j$ is perpendicular to $\tn^j$ vector.
Due to the fact that $D_a \tB^a = 0$, one arrives at 
\ben
\int_{M_1} dV~P_a~\eta^a &=& 
{1 \over 8 \pi} \int_{\p M_1} d\Sigma~\bigg( K_{ij} - K_{a}^{a}~h_{ij} \bigg)~\eta^i~\tn^i 
+  {1 \over 4\pi}\sum\limits_{n=1}^N~\int_{\p M_1} d\Sigma~A_{k}^{(n)} \eta^k~\tn_i~ \tB^{i (n)} \\ \nonumber
&=& \tJ( \Sigma_2) - \tJ(\Sigma_1).
\een
If the left-hand side of the above relation is equal to zero the total angular momentum is conserved.
Moreover, it can be revealed, using the definition of $\tJ(\Sigma)$, that this quantity is invariant
with respect to the gauge transformation $A_i \rightarrow A_i + D_i \theta$. Of course one ought to assume that
$\theta$ disappear near infinity and equation (\ref{sym}) is fulfilled.
\par
Next we consider the behaviour of the total angular momentum near infinity. It suffices to
examine the expression given by
\be
\int_{S_{r \rightarrow \infty}} d\Sigma~\tB^{(n)}_i~\tn^i~A^{(n)}_k~\eta^k,
\ee
for each of the gauge field in the theory in question. We assume that $A^{(n)}_k \sim \cO(1/r)$,
$\tA^{(n)}_k \sim \cO(1/r)$ and for the Killing vector field one has that
$\mid \eta \mid \sim \mid x\p_y -y\p_x \mid = \cO(\rho)$. On the other hand, for the
{\it magnetic} and {\it electric} one suppose that they are proportional to $\p_r/r^2 + \cO(1/r^3)$. 
In Ref.\cite{dai13} the typical construction avoiding the difficulty of removing {Dirac string}
bounded with each asymptotical point $i_k$ was performed. 
One removes from the manifold in question the portion of the $z$-axis below or above
the adequate asymptotical point. The aforementioned method enables one to obtain $U(1)$ invariant
potential for each of the gauge field
$A^{(n)}_k$ in the form as
\be
A^{(n)}_i = {1 \over 2 k}~\sum\limits_{k=1}^M
\bigg( A^{(n)}_{+ i} + A^{(n)}_{- i} \bigg),
\ee
on $R^3  \setminus \{ z-axis \}$. 
Having in mind the asymptotical behaviours described above one can show that
\be
\lim_{r \rightarrow \infty}
{1 \over r^2}~
{1 \over 2 k}~\sum\limits_{k=1}^M~\sum\limits_{n=1}^N
 \bigg( A^{(n)}_{+ i} + A^{(n)}_{- i} \bigg)~\eta^i = 0.
\ee
Just the total angular momentum $\tJ$ tends at infinity to $J(\Sigma)$. Summing it 
all up, one can formulate the statement\\
Theorem:\\
Let $(M,~h_{ij},~K_{ab},~E_i,~\tE_i,~B_a,~\tB_a,~\la)$ be initial axisymmetric data of the quantities defined above. If
$P_k~\eta^k = 0$, then $\tJ$ is concerned, i.e., that for two $U(1)$ invariant hypersurfaces 
$\Sigma_1$ and $\Sigma_2$ and bounded domain one has that
\be
\tJ( \Sigma_1) = \tJ(\Sigma_2).
\ee
Moreover, $\tJ$ is invariant under the gauge transformations vanishing in the nearby of asymptotic 
regions, following, that
\be
\tJ(S_\infty) = J.
\ee
\par
By analogy with Einstein-Maxwell theory, we would like to pay some attention to the problem of the
so-called twist potential. It yields
\be
d \la = \ep_{abc}~\bigg( \pi^{bk} - 2~\sum\limits_{n=1}^N \theta^{(n)bk} \bigg)~\eta^c \eta_k~dx^a,
\ee
where
\be
\pi_{ab} = K_{ab} - K_{c}{}{}^{c}~h_{ab}, \qquad
\theta^{(n)}_{ab} = \ep_{imb}~\tB^{(n) i}~\ep_{a}{}{}^{lm}~A_l^{(n)}~\eta^b.
\ee
It will be interesting to elaborate conditions for which the twist potential exists. Namely, we
calculate $(d \la)_{ij}$. After using equation (\ref{div}) and properties of Killing vector fields
$\eta_\alpha$ it can be found that the following is fulfilled:
\ben
(d \la)_{ij} &=& D^a \bigg( \pi_{ab}~\eta^b - 2~\sum\limits_{n=1}^N \theta^{(n)}_{ab}~\eta^b
\bigg)~\ep_{ijl}~\eta^l 
\\ \nonumber
&=& \bigg( 8\pi~P_{j}~\eta^j + 2~\sum\limits_{n=1}^N ~D_i \tB^i~A_b~\eta^b \bigg)~\ep_{ijl}~\eta^l.
\een
From the above relation one has that if $P_a~\eta^a = 0$ the twist potential form is closed,
i.e.,  $(d \la)_{ij} = 0$. Moreover, as we assumed previously the manifold in question is
simply connected and all these facts imply that the twist potential exist.


\par
In \cite{gib06,chr08} it was shown that the ADM mass $m$ can be written in the form as
\be
m = {1 \over 16~\pi} \int dx^3
~\bigg[
{}^{(3)}R + {1 \over 2} ~\rho^2 ~e^{- 4 \alpha + 2 U}~\bigg(
\rho~W_{\rho,z} - W_{z, \rho} \bigg)^2 \bigg]~e^{2 \alpha - 2 U}
+ {1 \over 8 \pi}~\int dx^3~(D U)^2.
\label{mas}
\ee
Because of the fact that we consider a simply connected manifold, one enables  
to justify the existence of the potentials for each of the gauge field. This implies the following relations:
\ben
\na_\alpha \zeta^{(n)} &=& F_{\alpha \mu}^{(n)} \eta^{\alpha}, \qquad \na_\alpha \psi^{(n)}
= \ast F_{\alpha \mu}^{(n)} \eta^{\alpha},\\
\na_\alpha {\tilde \zeta}^{(n)} &=& \tF_{\alpha \mu}^{(n)} \eta^{\alpha}, \qquad \na_\alpha \tpsi^{(n)}
= \ast \tF_{\alpha \mu}^{(n)} \eta^{\alpha},
\een
In the orthonormal basis one has that
\be
\p_\alpha \Phi = \sqrt{g_{\varphi \varphi}}~\Gamma^{(n)}_{3 \alpha},
\ee
where $\Phi = (\zeta^{(n)},~{\tilde \zeta}^{(n)},~ \psi^{(n)},~\tpsi^{(n)})$
and $\Gamma^{(n)}_{3 \alpha} = (F_{\alpha \mu}^{(n)},~\ast F_{\alpha \mu}^{(n)},
~\tF_{\alpha \mu}^{(n)},~\tF_{\alpha \mu}^{(n)})$.
In Ref.\cite{dai08} it was found that the potential was bounded with the extrinsic curvature tensor 
$K_{ij}$ by the relation of the form
\be
\la = 2~\ep_{ijk}~K^j{}{}_l~\eta^k~\eta^l~dx^i.
\ee
On the other hand, one can find that the following equality is valid:
\be
e^{2 \alpha - 2U} ~\mid K \mid_{h}^2 
\ge 2~e^{2 \alpha - 2U}~(K_{13}^2 + K_{23}^2) 
= {e^{4U} \over 2~\rho^4}~\mid \la \mid_{h}^2.
\label{costa}
\ee
We also assume that the initial data set in maximal, i.e., $K_{j}{}{}^j = 0$.
Then, we insert equation (\ref{costa}) into relation (\ref{mas}). The outcome is provided by
\ben
m &\ge& {1 \over 16 \pi} \int dx^3~
\bigg[ {}^{(3)}R~e^{2 \alpha - 2U} + 2~(DU)^2 \bigg] \\
&\ge& {1 \over 16\pi} \int dx^3 ~\bigg[
(DU)^2 + {e^{4U} \over 2~\rho^4}~\mid \la \mid_{h}^2 + 2
~{e^{2U} \over \rho^2}
~\sum\limits_{n=1}^N
\bigg( D \psi^{(n)}~D {\tilde \zeta}^{(n)} - D \zeta^{(n)}~D \tpsi^{(n)} \bigg) \bigg].
\een
It happened that, the last term in the above inequality can be rearrange in the way given by
\be
B_i^{(n)}~\tE^{(n)i} - E_j^{(n)}~\tB^{(n)j} = e^{-2 \phi}~{e^{2U} \over \rho^2}~
\bigg[ (D \zeta^{(n)})^2 + (D \psi^{(n)} )^2 \bigg],
\ee
where we have taken into account the definition of $\tF_{ab}^{(n)}$.
Further one can define {\it dilaton-electric} and {\it dilaton-magnetic} charges 
and the adequate potentials for each of the gauge fields under considerations.
Namely, they are provided by the relations of the forms as
\ben
\na_{\alpha} {\hat \psi}^{(n)} &=& e^{- \phi}~\ast F_{\beta \alpha}^{(n)}~\eta^\beta, \qquad 
Q_{d-e}^{(n)} = - {1 \over 4 \pi} \int dS_{\alpha \beta} ~e^{- \phi}~\ast F^{\alpha \beta (n)}
= {{\hat \psi}^{(n)}_2 - {\hat \psi}^{(n)}_1 \over 2},\\
\na_{\alpha} {\hat \zeta}^{(n)} &=& e^{- \phi}~F_{\beta \alpha}^{(n)}~\eta^\beta, \qquad 
Q_{d-m}^{(n)} = - {1 \over 4 \pi} \int dS_{\alpha \beta} ~e^{- \phi}~F^{\alpha \beta (n)}
= {{\hat \zeta}^{(n)}_1 - {\hat \zeta}^{(n)}_1 \over 2}.
\een
By virtue of the above definitions
we can invoke all the procedure elaborated in Refs.\cite{chr08}-\cite{lop10}, to find an inequality
binding the black hole ADM mass with other quantities characterizing black hole in EMAD-gravity.
Namely, one defines action
\be
I = \int dx^3~\bigg[
(D U)^2 + {e^{4 U} \over 2 \rho^4} \mid \la \mid_h^2 + 2~{e^{2U} \over \rho^2} ~\sum\limits_{n=1}^N
\bigg( (D {\hat \zeta}^{(n)})^2 + (D {\hat \psi}^{(n)})^2 \bigg) \bigg].
\ee
Then we use harmonic map associated with the extreme Kerr-Sen solution 
$I({\tilde \mu},~{\tilde \la},~{\tilde {\hat \zeta}},~{\tilde {\hat \psi}})$. One would like to show that
$I({\tilde \mu},~{\tilde \la},~{\tilde {\hat \zeta}},~{\tilde {\hat \psi}}) \ge
I({\mu},~{\la},~{\hat \zeta},~{\hat \psi})$. It can be shown by the methods developed in Refs.
\cite{chr08}-\cite{lop10}, so we refer
readers to the original works for particulars. Consequently, we can conclude that\\
Theorem:\\
Let $(M,~h_{ij},~K_{ij},~{\hat \zeta},~{\hat \psi},~\la)$ be a smooth three-dimensional maximal
time symmetric data set on simply
connected manifold which is invariant under the action of $U(1)$-group, with two asymptotically flat regions. 
Suppose further that there is no electromagnetic matter fields and the dominant energy condition is
assured. Then the ADM mass $m$, angular momentum and global {\it dilaton-electric} and 
{\it dilaton-magnetic} charges 
one obtains the inequality 
\be
m \ge \sqrt{{ \mid {\vec J} \mid^2 \over m^2} + 
\sum\limits_{n=1}^N ~\bigg(
{Q_{d-e}^{(n)}}^ 2 + {Q_{d-m}^{(n)}}^ 2 \bigg)}.
\label{mass}
\ee
By the direct calculations one can readily verify that the above inequality can be rewritten in the
analogous form
\be
m^2 \ge {Q_{(N)}^2 + \sqrt{ {Q_{(N)}^4 + 4~\mid {\vec J} \mid^2}} \over 2},
\ee
where for the brevity of notation we set $Q^2_{(N)}$ equal to
\be
Q^2_{(N)} = \sum\limits_{n=1}^N \bigg(
{Q_{d-e}^{(n)}}^ 2 + {Q_{d-m}^{(n)}}^ 2 \bigg).
\ee

\section{Area inequalities}
In this section we comment on
the inequality connecting the area, angular momentum and charges for
dynamical black holes in EMAD-gravity. To commence with, one considers a closed orientable two-dimensional
spacelike surface $\cS$ smoothly embedded in the manifold in question. 
Its intrinsic geometry is characterized by induced metric $q_{ab}$ with connection ${}^{(2)}D_a$,
Ricci scalar ${}^{(2)}R$, volume element $\ep_{ab}$ and area measure $d \cS$. As far as the extrinsic geometry
is concerned one introduces the normal outgoing and ingoing null vectors $l^i$ and $k^j$ normalized as
$l^i~k_i = -1$. Therefore the extrinsic geometry is characterized by the expansion $\theta^{(l)}$,
the shear $\si^{(l)}_{ij}$ and the normal fundamental form 
$\Omega^{(l)}_j$ bounded with the outgoing normal null vector $l^a$.
They are provided by the relations as follows:
\ben
\theta^{(l)} &=& q^{ab}~\na_{a}l_b, \qquad
\si^{(l)}_{ij} = q^{c}{}{}_i~q^d{}{}_j~\na{c} l_d - {1 \over 2}~\theta^{(l)}~q_{ij},\\ \nonumber
\Omega^{(l)}_j &=& - k^m~q^{r}{}{}_j~\na_m l_r.
\een
Moreover we require that the surface $\cS$ is the marginally outer trapped surface, i.e., $\theta^{(l)} = 0$,
as well as we demand that the hypersurface in question is stable. It means that there is an outgoing
vector $X^a = \la_1~l_a - \la_2~k_a$, with $\la_1 \ge 0$ and $\la_2 > 0$ satisfying
the condition of the form as 
$\delta_X \theta^{(l)} \ge 0$. The operator $\delta_X$ is the variation operator on surface $\cS$ along
the vector $X^a$ \cite{jar11}. Additionally
the surface should be axisymmetric with the Killing vector field $\eta_a$ and the following relations
should be given
\be
\cL_\eta l^j = \cL_\eta k^j = \cL_\eta \Omega^{(l)}_j = \cL_\eta F^{(n)} = \cL_\eta \la = 0,
\ee
where $F^{(n)}$ denotes the projection of the adequate strength of $n$-th gauge field. \\
In Ref.\cite{jar11} it was revealed that for a closed marginally trapped surface $\cS$ satisfying the
stably outermost condition for vector $X^a$ and for every axisymmetric function $\alpha$,
the following inequality implied
\ben
\int_{\cS} d\cS ~&\bigg(& {}^{(2)}D_a \alpha~{}^{(2)}D^a \alpha + {1 \over 2}
\alpha^2~{}^{(2)}R \bigg) \ge
\int_{\cS} d\cS ~\bigg( \alpha^2~\Omega^{(\eta)}_j\Omega^{(\eta) j}
+ \alpha~\beta~\si^{(l)}_{ij}~\si^{(l) ij} \\ \nonumber
&+& G_{ab}~\alpha~l^a~(\alpha ~k^b + \beta~l^b) \bigg),
\een
where $\beta = \alpha~\la_1/\la_2$.
In the case of EMAD-gravity the right-hand side of the above inequality is equal to relation
\ben \label{ineq}
\int_{\cS} d\cS ~&\bigg[&
\alpha^2~\Omega^{(\eta)}_j\Omega^{(\eta) j}
+ \alpha \beta~\si^{(l)}_{ij}~\si^{(l) ij}
+ 2~ \alpha \beta~(l^j \na_j \phi)^2
+ {1 \over 2}e^{4 \phi}~\alpha^2~(l^j \na_j a)^2 \\ \nonumber
&+&
2~\alpha \beta~\sum\limits_{n=1}^N (F_{ak}^{}{}^{(n)}~l^a)~(F_{j}{}{}^{(n) k}~l^j)
+ T_{ij}(matter)\alpha l^i~(\alpha ~k^j + \beta~l^j) \\ \nonumber
&+& \alpha^2~\bigg(\Omega^{(\eta)}_j\Omega^{(\eta) j}
+ {}^{(2)}D_i \phi {}^{(2)}D^i \phi + {1 \over 4}~e^{4 \phi}~{}^{(2)}D_k a~{}^{(2)}D^k a
+
e^{- 2 \phi} ~\sum\limits_{n=1}^N ( \cE^2 + \cB^2) \bigg)\bigg],
\een
where $\cE = F_{ab}~l^a~k^b$ and $\cB = \ast F_{ij}~l^i~k^j$. 
Because of the fact 
that we assume the dominant energy condition for matter fields, i.e.,
$T_{ij}(matter)\alpha l^i~(\alpha ~k^j + \beta~l^j) \ge 0$ as well as null energy condition for
$U(1)$-gauge fields $2~\alpha \beta~\sum\limits_{n=1}^N (F_{ak}^{}{}^{(n)}~l^a)~(F_{j}{}{}^{(n) k}~l^j)
\ge 0$, one obtains all positive terms on the right-hand side of equation (\ref{ineq}).
Abandoning the non-negative terms
we arrive at the following relation:
\ben
\int_{\cS} d\cS ~&\bigg(& {}^{(2)}D_a \alpha~{}^{(2)}D^a \alpha + {1 \over 2}
\alpha^2~{}^{(2)}R \bigg) \ge \\ \nonumber
\int_{\cS} d\cS ~\alpha^2~&\bigg[&
\Omega^{(\eta)}_j\Omega^{(\eta) j} + 
{}^{(2)}D_i \phi {}^{(2)}D^i \phi + {1 \over 4}~e^{4 \phi}~{}^{(2)}D_k a~{}^{(2)}D^k a
+
e^{- 2 \phi} ~\sum\limits_{n=1}^N ( \cE^2 + \cB^2) \bigg)\bigg].
\een
To have a closer insight in the inequality we introduce the following
axisymmetric line element on the two-dimensional surface $\cS$
\be
ds^2 = q_{ab}~dx^a~dx^b = e^{\si}~ \bigg( e^{2 q}~d \theta^2 + \sin^2 \theta~d \varphi^2 \bigg),
\ee  
where $\si + q = const = c$. 
Now it should be recalled \cite{jar11} that the fundamental form $\Omega^{(l)}_a$ can be festered
by means of the Hodge decomposition, i.e., $\Omega^{(l)}_a = \ep_{ab}~D^b \omega + D_a \tla$.
Because of the fact that $\Omega^{(l)}_a$ is axisymmetric one can readily verify that
$\Omega^{(\eta)}_a = {1 \over 2 \eta}~\ep_{ab}~D^b \omega$. It was revealed in the preceding section that
the total angular momentum consist of the gravitational part and the element contributed to the gauge fields in the
underlying theory. The gravitational branch of the angular momentum is given by
\be
J = {1 \over 8\pi}\int_{\cS} dS~\Omega^{(l)}_a~\eta^a = {\omega(\pi) - \omega(0) \over 8}.
\ee
In order to describe the other part of the potential it is useful to introduce
another potential \cite{gal95} of the form
\be
d \chi = \sum\limits_{n=1}^N~\bigg( 2~\eta~d \omega -
2 v^{(n)}~d k^{(n)} + 2~k^{(n)}~d v^{(n)} \bigg).
\ee
Then, by a direct calculation it can be revealed that
$d\cS = e^c~dS_0$, where $ dS_0 = \sin \theta d\theta d \varphi$. 
Choosing $\alpha = e^{c - \si/2}$
one achieves at
\ben \label{m}
2(c+1) &\ge& {1 \over 2 \pi}~\int_{\cS} dS_0~\bigg[
\si + {1 \over 4}~D_{m} \si D^M \si
+  D_{a} \phi D^a \phi 
+ {1 \over 4}~e^{4 \phi}~D_{i} a D^i a \\ \nonumber
&+& 
{1 \over 4 \eta^2}~\mid
\sum\limits_{n=1}^N~\bigg( 
D_j\chi + 2 v^{(n)}~D_j k^{(n)} - 2~k^{(n)}~D_j v^{(n)}
\bigg) \mid^2 \\ \nonumber
&+& {1 \over \eta}~\sum\limits_{n=1}^N~D_{a}  {\hat \psi}^{(n)} D^a {\hat \psi}^{(n)}
+ {1 \over \eta}~\sum\limits_{n=1}^N~D_{b}  {\hat \zeta}^{(n)} D^b {\hat \zeta}^{(n)}
\bigg],
\een
where $\eta = q_{ij} \eta^i \eta^j$.
Having in mind that $A = 4\pi~e^c$ one can reach the inequality
\be
A \ge 4\pi~e^{\cM -2 \over 2},
\ee
where the functional $\cM$ is defined as the right-hand side of the equation (\ref{m}).
\par
In order to the inequality among area, angular momentum and charges is to utilize the connection between
the functional $\cM$ and a harmonic energy for maps from the sphere into the complex hyperbolic space. 
The key point in the proof is to show that the extreme Kerr-Sen sphere, i.e., the set fulfilling
the Lagrange equations for the functional $\cM$ \cite{ace11}.
\par
One should also take into account the results of Ref.\cite{hil77}, which state that if the domain for the map
is compact, connected, with
non-void boundary and the target manifold has negative sectional curvature, then the minimizer
of the harmonic energy subject to the Dirichlet boundary conditions, exists.
Just, on account of this result one can conclude that harmonic maps are minimizers of the harmonic energy
for the given Dirichlet boundary conditions.\\
As in Ref.\cite{cle12} we relate the functional $\cM$ to the standard harmonic energy
$\cM_D$ from a subset $D \subset S^2 \setminus \{\theta = 0,~\pi \}$ to the complex hyperbolic space
with the strictly positive line element provided by
\be
ds^2_H = {d \eta^2 \over \eta}
+ {1 \over \eta^2}~{\bigg[ \sum\limits_{n=1}^N~( d \chi + 2 v^{(n)}~dk^{(n)} - 2~k^{(n)}~dv^{(n)}) \bigg]^2}
+ {1 \over \eta}~\sum\limits_{n=1}^N~\bigg[ (d {\hat \zeta}^{(n)})^2 + (d {\hat \psi}^{(n)})^2 \bigg],
\ee
while $\cM_D$ implies 
\be
\cM_D = \cM + \int_{\cS} dS~\ln \sin \theta + \int_{\p \cS}dl~(\si + \ln \sin \theta)~{\p \ln \sin \theta
\over \p n}.
\ee
We have set $n$ to be unit normal to the boundary to $\cS$ surface, while $dl$ is the measure
element of the boundary $\p \cS$. It should be noticed that both functionals have the same forms of the Lagrange 
equations because
of the fact that the difference between them is equal to a constant plus a boundary term. The proof
goes like in Ref.\cite{cle12} so we refer the reader for the mathematical details to the article.
For the convenience of the reader we sketch the main steps of it. Namely, we divide
the underlying sphere into three regions
\be 
\Omega_I = \{ \sin \theta \le e^{-(ln \ep)^2} \}, \qquad
\Omega_{II} = \{ e^{-(ln \ep)^2} \le \sin \theta \le \ep \}, \qquad 
\Omega_{III} = \{ \ep \le \sin \theta \}.
\ee
Firstly, we interpolate the potentials between extreme Kerr-Sen solution in $\Omega_I$ and
a general solution in $\Omega_{III}$ region.
It leads to the Dirichlet problem in $\Omega_{IV} = \Omega_{II} \cup \Omega_{III}$ which
yields that the mass functional of Kerr-Sen extreme solution is less than or equal to the mass
functional for the auxiliary interpolating map on the whole sphere. In the last step one has in mind 
the limit as $\Omega_{III}$ converges to the sphere and reveals that the mass functional in question 
for the auxiliary maps converges to the mass functional for the original sets. All these
mathematical machinery leads to the inequality
\be
e^{\cM -2} \ge 4~J^2 + Q_{(N)}^4.
\ee
One can make use of the above inequality. In the case when we consider two asymptotically flat ends
there exist an asymptotic stable (i.e., the second variation of the area is nonnegative) minimal surface
$\Sigma_{min} \in \cM$  which separates the aforementioned asymptotically
flat ends. As was remarked in \cite{dai13} $\Sigma_{min}$ minimizes area among all the considered two-surfaces
$A(\Sigma_{min}) = A_{min}$, where $ A_{min}$ is the least area pone requires to enclose the ends. Just having in mind
conclusions presented in Ref.\cite{dai13} and our inequality one obtains
\be
A_{min} \ge 4 \pi ~\sqrt{4~J^2(\Sigma) + Q^4_{(N)}(\Sigma)},
\label{area}
\ee
where in the above inequality $\Sigma$ stands either for $\Sigma_{min}$ or for $\Sigma_0$. The equality is satisfied when
$\Sigma = \Sigma_{0}$ and this case is responsible for the Kerr-Sen extreme sphere.
Using the relations (\ref{mass}) and (\ref{area}), one can conclude\\
Theorem:\\
Assume that one has axially symmetric, maximal and simply connected initial data set with two asymptotically flat
ends. Suppose moreover that we consider non-electromagnetic matter field in EMAD-gravity. Let us demand that the 
dominant energy condition and $P_k~\eta^k = 0$ is fulfilled, where $\eta_k$ is the axially symmetric Killing vector field.
Then, the following inequality is provided: 
\be 
{A_{min} \over 8 \pi} \ge m^2 - { Q_{(N)}^2 \over 2}
- \sqrt{\bigg( m^2 - { Q_{(N)}^2 \over 2} \bigg)^2 - { Q_{(N)}^4 \over 4}  - J^2},
\ee
where $A_{min}$ is the minimum area to enclose the asymptotically flat ends. The Kerr-Sen extremal spacetime 
is subject to the equality.

\section{Conclusions}
In our paper we have considered the EMAD-gravity theory being the low-energy limit of the heterotic string 
theory with arbitrary number of $U(1)$-gauge fields. 
Matter fields were assumed to be non-electromagnetic and satisfying the dominant energy condition.
We define the general form of the total angular momentum as well as the twist 
potential in the theory under consideration. Considering a smooth three-dimensional time
symmetric data set on simply connected manifold invariant under the action
of $U(1)$ group, with two asymptotically flat ends, we arrive at the inequality binding the ADM mass
angular momentum and global {\it dilaton-electric} and {\it dilaton-magnetic} charges.
Then, we examine a closed orientable two-dimensional spacelike surface which is smoothly embedded
in spacetime of EMAD-gravity. 
Then it was shown that the ADM mass is subject to inequality expressed in terms 
of the area angular momentum and charges of black holes in EMAD-gravity.
\par
Considering axially symmetric maximal and simply connected data set with two-asymptotically flat ends
and demanding the dominant energy condition and relation $P^a~\eta_a = 0$, we achieved at the inequality
expressing the area in terms of the ADM mass, angular momentum and charges in the underlying theory.
The inequality was saturated if the initial data emerge from the extremal Kerr-Sen black hole.


\begin{acknowledgments}
We thank Jose Luis Jaramillo for the fruitfull discussions on various ocassions.
MR was partially supported by the grant of the National Science Center
$2011/01/B/ST2/00488$.
\end{acknowledgments}




\begin{thebibliography}{99}
%
\def\cmp#1#2#3{{ Commun. Math. Phys.} {\bf #1}, #2 (#3)}
\def\lmp#1#2#3{{ Lett. Math. Phys.} {\bf #1}, #2 (#3)}
\def\cpam#1#2#3{{ Commun. Pure Appl. Math.} {\bf #1}, #2 (#3)}
\def\hpa#1#2#3{{ Hell. Phys. Acta} {\bf #1}, #2 (#3)}
\def\grg#1#2#3{{ Gen. Rel. Grav.} {\bf #1}, #2 (#3)}
\def\pr#1#2#3{{ Phys. Rev.} {\bf #1}, #2 (#3)}
\def\prl#1#2#3{{ Phys. Rev. Lett.} {\bf #1}, #2 (#3)}
\def\prd#1#2#3{{ Phys. Rev. D} {\bf #1}, #2 (#3)}
\def\pl#1#2#3{{ Phys. Lett} {\bf #1}, #2 (#3)}
\def\pla#1#2#3{{ Phys. Lett. A} {\bf #1}, #2 (#3)}
\def\plb#1#2#3{{ Phys. Lett. B} {\bf #1}, #2 (#3)}
\def\prep#1#2#3{{ Phys. Reports} {\bf #1}, #2 (#3)}
\def\phys#1#2#3{{ Physica} {\bf #1}, #2 (#3)}
\def\jcp#1#2#3{{ J. Comput. Phys.} {\bf #1}, #2 (#3)}
\def\jmp#1#2#3{{ J. Math. Phys.} {\bf #1}, #2 (#3)}
\def\jpm#1#2#3{{ J. Phys. A: Math. Gen.} {\bf #1}, #2 (#3)}
\def\cpr#1#2#3{{ Computer Phys. Rept.} {\bf #1}, #2 (#3)}
\def\cqg#1#2#3{{ Class. Quantum Grav.} {\bf #1}, #2 (#3)}
\def\cma#1#2#3{{ Computers Math. Applic.} {\bf #1}, #2 (#3)}
\def\mc#1#2#3{{ Math. Compt.} {\bf #1}, #2 (#3)}
\def\apj#1#2#3{{ Astrophys. J.} {\bf #1}, #2 (#3)}
\def\apjs#1#2#3{{ Astrophys. J. Suppl.} {\bf #1}, #2 (#3)}
\def\acta#1#2#3{{ Acta Astronomica} {\bf #1}, #2 (#3)}
\def\apl#1#2#3{{Ann. Physik. (Leipzig)} {\bf #1}, #2 (#3)}
\def\anp#1#2#3{{Ann. Phys. } {\bf #1}, #2 (#3)}
\def\ahp#1#2#3{{Ann. Henri Poincare } {\bf #1}, #2 (#3)}
\def\am#1#2#3{{Acta Math.} {\bf #1}, #2 (#3)}
\def\sa#1#2#3{{ Sov. Astro.} {\bf #1}, #2 (#3)}
\def\sia#1#2#3{{ SIAM J. Sci. Statist. Comput.} {\bf #1}, #2 (#3)}
\def\aa#1#2#3{{ Astron. Astrophys.} {\bf #1}, #2 (#3)}
\def\mnras#1#2#3{{ Mon. Not. R. astr. Soc.} {\bf #1}, #2 (#3)}
\def\npb#1#2#3{{ Nucl. Phys. B} {\bf #1}, #2 (#3)}
\def\prsla#1#2#3{{ Proc. R. Soc. London, Ser. A} {\bf #1}, #2 (#3)}
\def\jhep#1#2#3{{ JHEP} {\bf #1}, #2 (#3)}             
\def\nuc#1#2#3{{Nuovo Cimento B } {\bf #1}, #2 (#3)}
\def\ijmp#1#2#3{{Int. J. Mod. Phys. D} {\bf #1}, #2 (#3)}
\def\atmp#1#2#3{{Adv. Theor. Math. Phys.} {\bf #1}, #2 (#3)}
\def\ptps#1#2#3{{Prog. Theor. Phys. Suppl.} {\bf #1}, #2 (#3)}
\def\lmp#1#2#3{{Lett. Math. Phys. } {\bf #1}, #2 (#3)}
\def\mmj#1#2#3{{Mich. Math. j. } {\bf #1}, #2 (#3)}
\def\annny#1#2#3{{Ann.N.Y.Acad.Sci.} {\bf #1}, #2 (#3)}
\def\jdg#1#2#3{{ J. Diff. Geom.} {\bf #1}, #2 (#3)}
\def\hepph#1#2{{ hep-ph }{\bf #1} (#2)}
\def\hepth#1#2{{ hep-th }{\bf #1} (#2)}
\def\grqc#1#2{{ gr-qc }{\bf #1} (#2)}
\def\ibid#1#2#3{{ {\it ibid.} }{\bf #1}, #2 (#3)}
%
\bibitem{pen73}
R.Penrose, \annny{224}{125}{1973}.

\bibitem{bri59}
D.R.Brill, \anp{7}{466}{1959}.
\bibitem{gib06}
G.W.Gibbons and G.Holzegel, \cqg{23}{6459}{2006}.


\bibitem{dai06}
S.Dain, \prl{96}{101101}{2006},\\
S.Dain, \cqg{23}{6845}{2006}.
\bibitem{dai08}
S.Dain, \jdg{79}{33}{2008}.
\bibitem{chr08}
P.T.Chrusciel, \anp{323}{2566}{2008}.
\bibitem{chr08a}
P.T.Chrusciel, Y.Li, and G.Weinstein, \anp{323}{2591}{2008}.
\bibitem{chr09}
P.T.Chrusciel and J.Lopes, \cqg{26}{235013}{2009}.
\bibitem{lop10}
J.Lopes, \jpm{43}{285202}{2010}.

\bibitem{ansorg}
M.Ansorg and H.Pfister, \cqg{25}{035009}{2008},\\
J.Hennig, M.Ansorg, and C.Cederbaum,\ibid{25}{162002}{2008},\\
J.Hennig, M.Ansorg, and C.Cederbaum,\cmp{293}{449}{2010}.
\bibitem{dai10}
S.Dain, \prd{82}{104010}{2010}. 
\bibitem{ace11}
A.Acena, S.Dain, and M.E.Gabach Clement, \cqg{28}{105014}{2011}.
\bibitem{dai11}
S.Dain and M.Reiris, \prl{107}{051101}{2011}.
\bibitem{jar11}
J.L.Jaramillo, M.Reiris, and S.Dain, \prd{84}{121503(R)}{2011},\\
S.Dain, J.L.Jaramillo, and M.Reiris, \cqg{29}{035013}{2012},\\
W.Simon, \cqg{29}{062001}{2012}.
\bibitem{gab12}
M.E.Gabach Clement and J.L.Jaramillo, \prd{86}{064021}{2012}.
\bibitem{cle12}
M.E.Gabach Clement, J.L.Jaramillo, and M.Reiris, \grqc{1207.6761}{2012}.
\bibitem{dai12}
S.Dain, \cqg{29}{073001}{2012}.
\bibitem{mar12}
M.Mars, \cqg{29}{145019}{2012},\\
J.L.Jaramillo, \ibid{29}{177001}{2012}.
\bibitem{dai13}
S.Dain, M.Khuri, G.Weinstein, and S.Yamada, \prd{88}{024048}{2013}.














\bibitem{nd}
G.W.Gibbons, D.Ida, and T.Shiromizu, \prd{66}{044010}{2002},\\
G.W.Gibbons, D.Ida, and T.Shiromizu, \prl{89}{041101}{2002},\\
S.Hollands, A.Ishibashi, and R.M.Wald, \cmp{271}{699}{2007},\\
M.Rogatko, \cqg{19}{L151}{2002},\\
M.Rogatko, \prd{67}{084025}{2003},\\
M.Rogatko, \ibid{70}{084025}{2004},\\
M.Rogatko, \ibid{73}{124027}{2006}.

\bibitem{nrot}
Y.Morisawa and D.Ida, \prd{69}{124005}{2004},\\
Y.Morisawa, S.Tomizawa, and Y.Yasui, \prd{77}{064019}{2008},\\
M.Rogatko, \prd{77}{124037}{2008},\\
S.Hollands and S.Yazadjiev, \cmp{283}{749}{2008},\\
S.Hollands and S.Yazadjiev, \cqg{25}{095010}{2008},\\
D.Ida, A.Ishibashi, and T.Shiromizu, \ptps{189}{52}{2011}.

\bibitem{sugra}
A.K.M.Massod-ul-Alam, \cqg{14}{2649}{1993},\\
M.Mars and W.Simon, \atmp{6}{279}{2003},\\
M.Rogatko, \cqg{14}{2425}{1997},\\
M.Rogatko, \prd{58}{044011}{1998},\\
M.Rogatko, \ibid{59}{104010}{1999},\\
M.Rogatko, \ibid{82}{044017}{2010},\\
M.Rogatko, \cqg{19}{875}{2002},\\
S.Tomizawa, Y.Yasui, and A.Ishibashi, \prd{79}{124023}{2009},\\
S.Tomizawa, Y.Yasui, and A.Ishibashi, \prd{81}{084037}{2010},\\
J.B.Gutowski, \jhep{0408}{049}{2004},\\
J.P.Gauntlett, J.B.Gutowski, C.M.Hull, S.Pakis, and H.S.Real, \cqg{20}{4587}{2003}.


\bibitem{shi13a}
T.Shiromizu and S.Ohashi, \prd{87}{087501}{2013}.

\bibitem{shi12}
T.Shiromizu, S.Ohashi, and R.Suzuki, \prd{86}{064041}{2012}.

\bibitem{bak13}
B.Bakon and M.Rogatko, \prd{87}{084065}{2013}.

\bibitem{rog13}
M.Rogatko, \prd{88}{024051}{2013}.

\bibitem{hol12}
S.Hollands, \cqg{29}{065006}{2012}.

\bibitem{yaz13}
S.Yazadjiev, \prd{87}{024016}{2013}.
\bibitem{yaz13b}
S.Yazadjiev, \cqg{30}{115010}{2013}.
\bibitem{faj13}
D.Fajman and W.Simon, \grqc{1308.3659}{2013},\\
T.Torben Paetz and W.Simon, \cqg{30}{235005}{2013}.



\bibitem{kal}
E.Bergshoeff, R.Kallosh, and T.Ortin, \npb{478}{156}{1996},\\
R.Kallosh, A.Linde, T.Ortin, A.Peet, and A.Van Proeyen, \prd{46}{5278}{1992},\\
R.Kallosh and T.Ortin, \ibid{48}{742}{1993},\\
T.Ortin, \ibid{47}{3136}{1993},\\
R.Kallosh, D.Kastor, T.Ortin, and T.Torma, \ibid{50}{6374}{1994}.

\bibitem{gal95}
D.Gal'tsov, A.Garcia, and O.Kechkin, \jmp{36}{5023}{1995}.

\bibitem{hil77}
S.Hildebrandt, H.Kaul, and K.Widman, \am{138}{1}{1977}.

\end{thebibliography}
\end{document}